\documentclass{PoS}

\newcommand{\AGeV}[1][ ]{$A$~GeV{#1}}

\newcommand{\T}{$T$}
\newcommand{\muB}{$\mu_B$}

\title{Statistical Model and the mesonic-baryonic transition region
}

\ShortTitle{Statistical Model and the mesonic-baryonic transition
region}

\author{\speaker{H. Oeschler}\\
        Darmstadt University of Technology,
D-64289 Darmstadt, Germany\\
        E-mail: \email{h.oeschler@gsi.de}}

\author{J. Cleymans\\
UCT-CERN Research Centre and Department  of  Physics,\\
University of Cape Town, Rondebosch 7701, South Africa}
\author{K. Redlich\\
Institute of Theoretical Physics, University of Wroc\l aw,
Pl-45204 Wroc\l aw, Poland}
\author{S. Wheaton\\
UCT-CERN Research Centre and Department  of  Physics,\\
University of Cape Town, Rondebosch 7701, South Africa}

\abstract{The statistical model assuming chemical equilibrium and
local strangeness conservation describes most of the observed
features of strange particle production from SIS up to RHIC.
Deviations are found as the maximum in the measured K$^+/\pi^+$
ratio is much sharper than in the model calculations. At the
incident energy of the maximum, the statistical model shows that
freeze out changes regime from one being dominated by baryons at
the lower energies toward one being dominated by mesons. It will
be shown how deviations from the usual freeze-out curve influence
the various particle ratios. Furthermore, other observables
exhibit also changes just in this energy regime.
}

\FullConference{5th International Workshop on Critical Point and Onset of Deconfinement - CPOD 2009,\\
         June 08 - 12 2009\\
         Brookhaven National Laboratory, Long Island, New York, USA}

\begin{document}

\section{Introduction}

\begin{figure}[h]
\begin{center}
\includegraphics*[width=10cm]{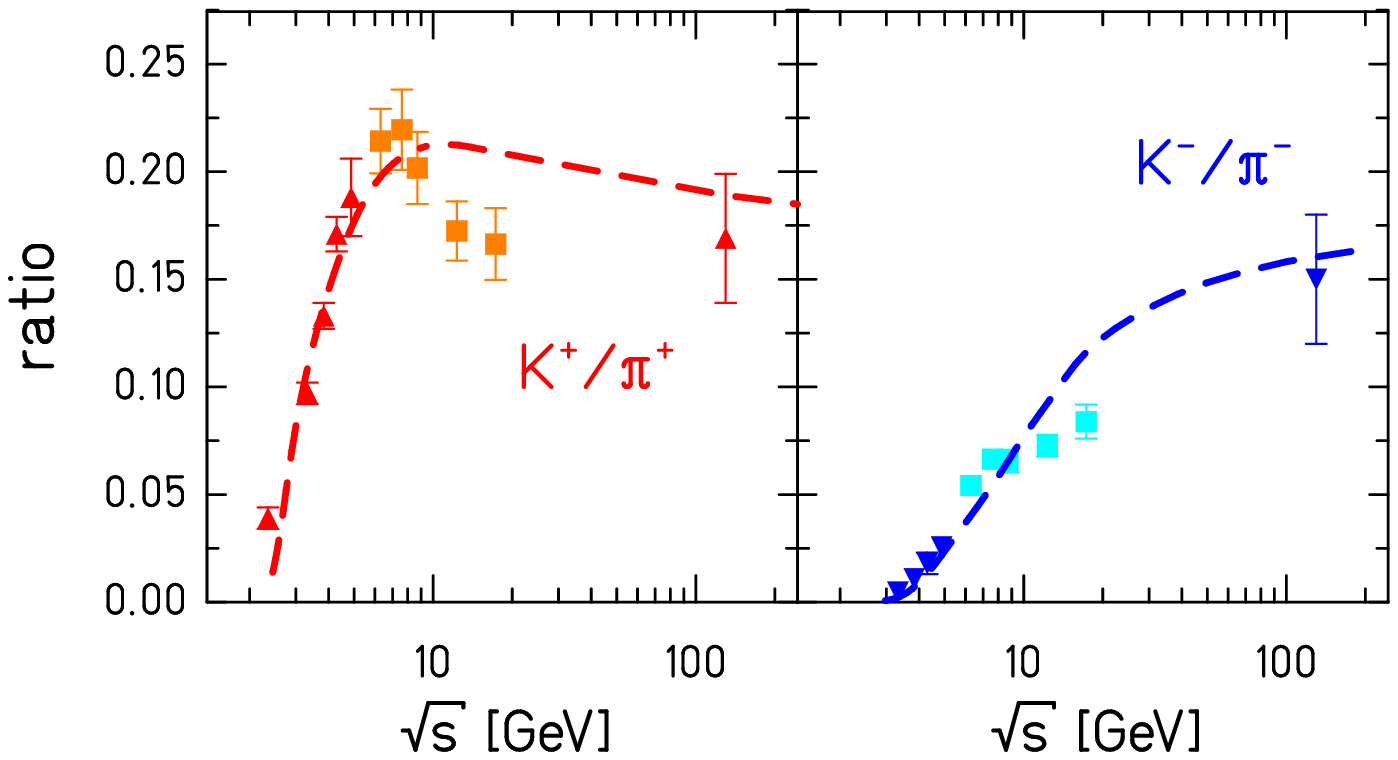}\\
\includegraphics*[width=13cm]{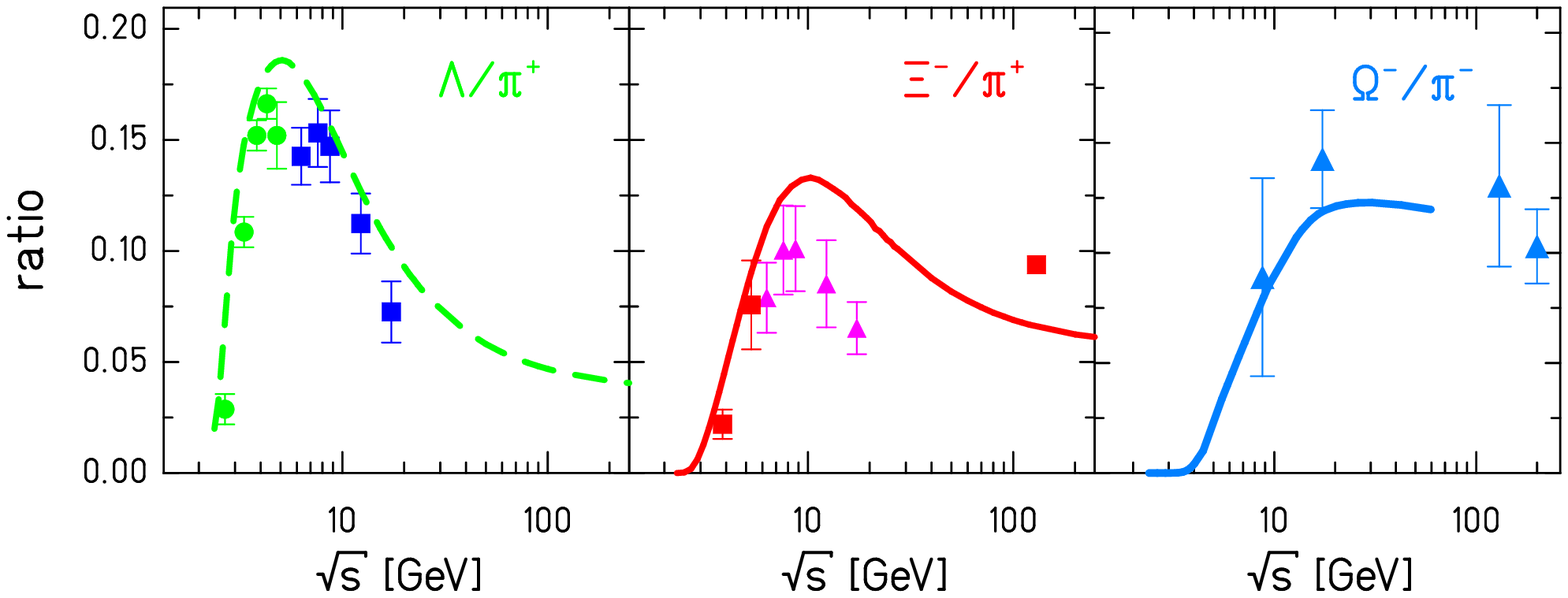}
 \caption{Ratio of strange-to-non-strange mesons (upper part) and
 the corresponding ratios for baryons (lower part) as a function of $\sqrt{s_{\rm NN}}$.}
\end{center}
\label{mesons_Baryons}
\end{figure}

The NA49 Collaboration  has performed a series of measurements of
the production of strange particles in central Pb-Pb collisions at
20, 30, 40, 80 and 158 $A$ GeV beam energies
\cite{NA49,Gazdzicki,Lambda-NA49,:2007fe,:2008vb}. The most
interesting result is the pronounced maximum in the K$^+/\pi^+$
ratio observed around 30 \AGeV. This ''horn'' has initiated a lot
of discussion  related as to whether or not it indicates a phase
transition. Indeed, this has been suggested
in~\cite{Gorenstein,Stock}. A more conventional interpretation has
been presented within the hadron gas model~\cite{max_strange}, yet
this model did not reproduce the measured yields in a satisfactory
manner. This description together with the recently published
values from NA49 and earlier AGS and RHIC results
\cite{pi-AGS,Ahle_1999,Ahle_2000,Lambda-AGS,Ahmad,Klay,Ahle,Dunlop,bearden,Star}
are summarized in Fig.~\ref{mesons_Baryons}.

It is important to remark that only the K$^+/\pi^+$ ratio exhibits
a sharp maximum while the K$^-/\pi^-$ ratio shows a continuous
rise with $\sqrt{s_{\rm NN}}$. The dashed and solid lines
represent calculation within the statistical
model~\cite{max_strange} explained in the next section. Both
trends are qualitatively described within this model.

In the lower part of Fig.~\ref{mesons_Baryons} the corresponding
ratios with strange baryons over pion are seen to exhibit also a
maximum, most pronounced for the ratio $\Lambda/\pi^+$. For the
other ratios the experimental situation is less clear and more
results are eagerly needed.

\section{Maximum relative strangeness content around $\sqrt{s_{\rm NN}}$ $\approx 6 -
8$ GeV}

\begin{figure}[h]
\begin{center}
\includegraphics*[width=7.5cm]{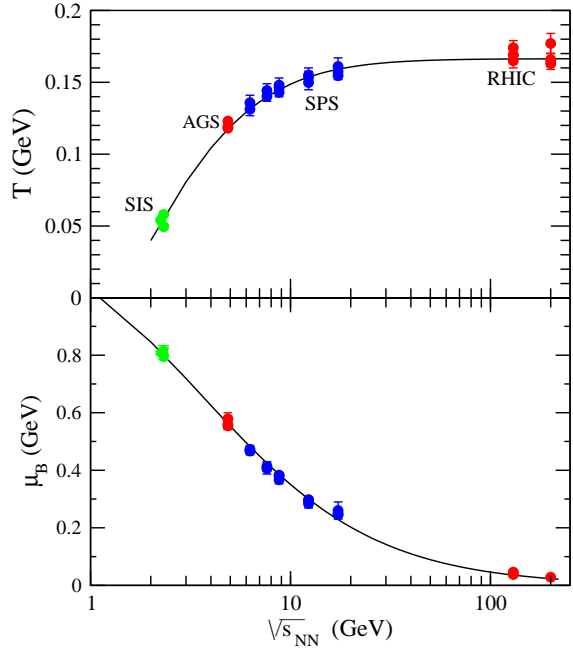}
\end{center}
\caption{Energy dependence of the chemical freeze-out parameters
$T$ and $\mu_B$. The curves have been obtained using a
parametrization discussed in the text. }
\label{muB_T_e}
\end{figure}

The statistical model is  very successful in describing particle
yields from SIS up to RHIC energies with only two parameters \T~
and \muB. At the very low incident energies a canonical
description with exact strangeness conservation is
needed~\cite{CLE99}. The extracted parameters \T~ and \muB~
plotted in a ''phase diagram'' describe a smooth line which can be
parameterized e.g.~by the $E/N \approx 1$ GeV condition
\cite{1gev}. Figure ~\ref{muB_T_e} shows these values as a
function of $\sqrt{s_{\rm NN}}$ exhibiting for \T~ a rising curve
which saturates above top SPS energies at a value of about 170
MeV.  The other parameter \muB~ decreases with incident energy
from a value near the nucleon mass to zero for fully transparent
collisions. The lines represent parameterizations
\begin{equation}
T(\mu_B) = a - b\mu_B^2 -c \mu_B^4 . \label{Eqn:T(muB)}
\end{equation}
where $ a =  0.166 \pm 0.002$ GeV, $b = 0.139 \pm 0.016$
GeV$^{-1}$ and $c = 0.053 \pm 0.021$ GeV$^{-3}$,
and

\begin{equation}
\mu_B(\sqrt{s}) = \frac{d}{1 + e\sqrt{s}}, \label{Eqn:MuB(s)}
\end{equation}
with $d = 1.308\pm 0.028$ GeV and  $e = 0.273 \pm 0.008$
GeV$^{-1}$.

Based on this set of the $\sqrt{s_{\rm NN}}$ dependence of the
thermal parameters the dashed lines in Fig.~\ref{mesons_Baryons}
have been calculated. They described the observed trends
qualitatively, but not the sharp maximum in K$^+/\pi^+$. Recently,
the statistical model has been extended including higher
resonances~\cite{Andronic:2008gu}. As they mostly decay into pions
the strong drop of the K$^+/\pi^+$ ratio towards RHIC energies as
observed in the data, is now much better described.

\begin{figure}[h]
\begin{center}
\includegraphics*[width=7.5cm]{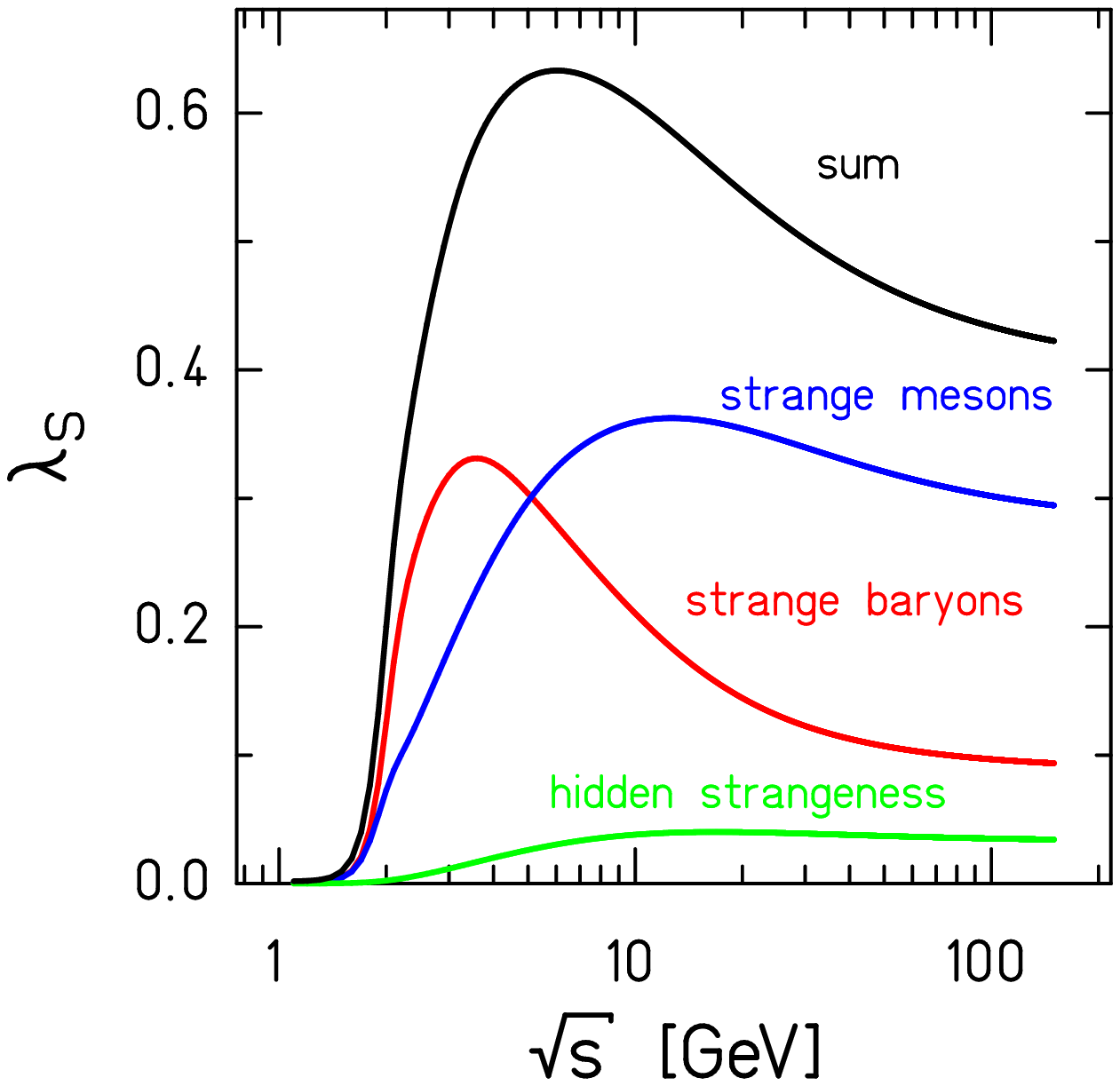}
\includegraphics*[width=7.5cm]{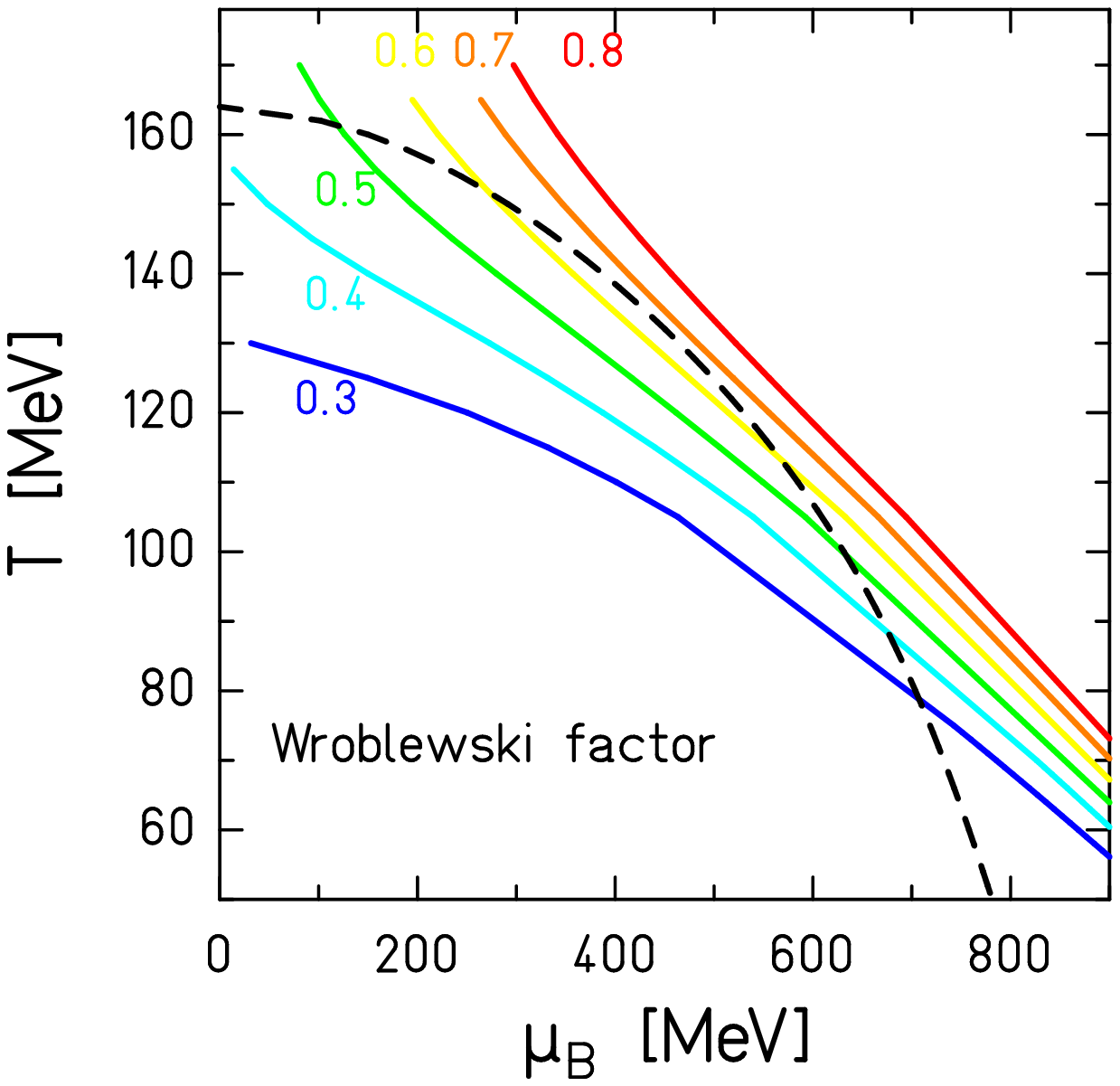}
\caption{Left: Contributions to the Wroblewski factor $\lambda_s$
(for definition see text) from strange baryons, strange mesons,
and mesons with hidden strangeness. The sum of all contributions
is given by the full black line. Right: Lines of constant
Wroblewski factor $\lambda_s$ in the $T-\mu_B$ plane (solid lines)
together with the freeze-out curve (dashed
line)~\protect\cite{1gev}.}
\end{center}
\label{Wrob}
\end{figure}

To study whether strangeness has a maximum or not, it is more
convenient to plot the Wroblewski factor~\cite{wroblewski} defined
as $$ \lambda_s \equiv {2\bigl<s\bar{s}\bigr>\over
\bigl<u\bar{u}\bigr> + \bigl<d\bar{d}\bigr>} $$ where the
quantities in angular brackets refer to the number of newly formed
quark-antiquark pairs, i.e.~$\lambda_s$ excludes all quarks that
were present in the target and the projectile nuclei. Figure~3,
left,  shows as solid line (marked ``sum'') the Statistical-Model
calculations along the unified freeze-out curve~\cite{1gev} with
the energy-dependent parameters $T$ and $\mu_B$ given above. From
this figure we conclude that around $\sqrt{s_{\rm NN}}$ = 6 GeV
corresponding to an incident energy of 20 $A$ GeV, the relative
strangeness content in heavy-ion collisions reaches a clear and
well pronounced maximum. The Wroblewski factor decreases towards
higher energies and reaches a limiting value of 0.43. For details
see Ref.~\cite{max_strange}.

The appearance of the maximum can be traced  to the specific
dependence of $\mu_B$ and $T$ on the beam energy as also pointed
out in Ref.~\cite{SK}. Figure~3, right, shows lines of constant
$\lambda_s$ in the $T-\mu_B$ plane. As expected, $\lambda_s$ rises
with increasing $T$ for fixed $\mu_B$. Following the chemical
freeze-out curve, shown as a dashed line in Fig.~\ref{Wrob}, one
can see that
 $\lambda_s$ rises quickly from SIS to AGS energies,
then reaches  a maximum at $\mu_B\approx 500$ MeV and $T\approx
130$ MeV. These freeze-out parameters correspond to 30 $A$ GeV
laboratory energy. At higher incident energies the increase in $T$
becomes negligible but $\mu_B$ keeps on decreasing and as a
consequence $\lambda_s$ also decreases.

The importance of finite baryon density on the behavior of
$\lambda_s$ is demonstrated in  Fig.~3, left, showing separately
the contributions to $\left<s\bar{s}\right>$ coming from strange
baryons, from strange mesons and from hidden strangeness,
i.e.~from hadrons  like $\phi$ and $\eta$. The origin of the
maximum in the Wroblewski ratio can be traced essentially to the
contribution of strange baryons. The production of strange baryons
dominates at low $\sqrt{s_{\rm NN}}$ and loses importance at high
incident energies when the yield of strange mesons increases.
However, strange mesons also exhibit a maximum, yet less
pronounced. This is due to the fact that strangeness production at
the lower energies occurs via the associated production,
i.e.~K$^+$ are created together with
hyperons~\cite{Cleymans:2004bf}. Therefore the K$^+$ mesons are
affected by the properties of the baryons, but the K$^-$ are not.

\section{Transition from Baryonic to mesonic freeze out}

While the Statistical Model cannot fully explain the sharpness of
the peak in the K$^+/\pi^+$ ratio, there are nevertheless several
phenomena giving rise to the rapid change which warrant a closer
look at the model.

\begin{figure}[h]
\begin{center}
\includegraphics[width=10.5cm]{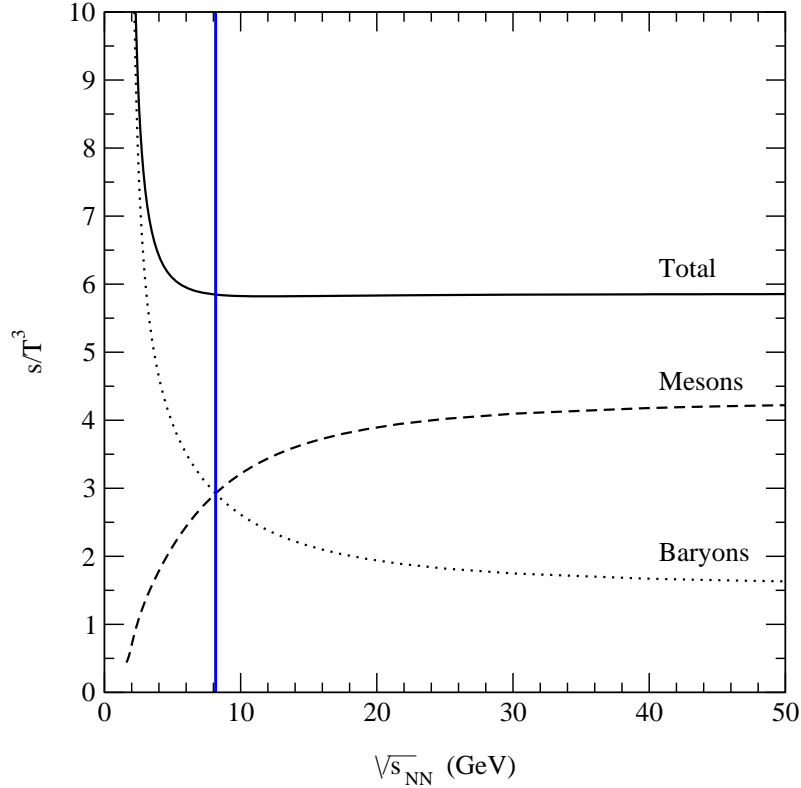}
 \caption{The entropy density normalized to $T^3$
as a function of the beam energy as calculated in the Statistical
Model using {\sc Thermus}~\cite{thermus}.} \end{center}
\label{entropy_s}
\end{figure}

It has been shown that $s/T^3$ = 6 is a fairly good criterium to
describe the freeze-out curve~\cite{Cle:2006} and we use it here
to describe the nature of the rapid change in the various ratios.
We show in Fig.~4 the entropy density divided by $T^3$ as a
function of beam energy as solid line.
The separate contribution of mesons and of baryons to the total
entropy is also shown in this figure by the dashed and dotted
lines. There is a clear change of baryon to meson dominance around
$\sqrt{s_{\rm NN}}$ = 8 GeV. Above this value the entropy is
carried mainly by mesonic degrees of freedom. It is remarkable
that the entropy density divided by $T^3$ is almost constant for
all incident energies  above the top AGS.

The separation line between meson dominated and baryon dominated
areas in the $T-\mu_B$ plane is given in Fig.~5. In this figure
the separation line crosses the freeze-out line at the stated
$\sqrt{s_{\rm NN}}$. This figure invites for further speculations
as e.g.~an existence of a triple point~\cite{triplepoint}.

\begin{figure}[h]
\begin{center}
\includegraphics[width=10.5cm]{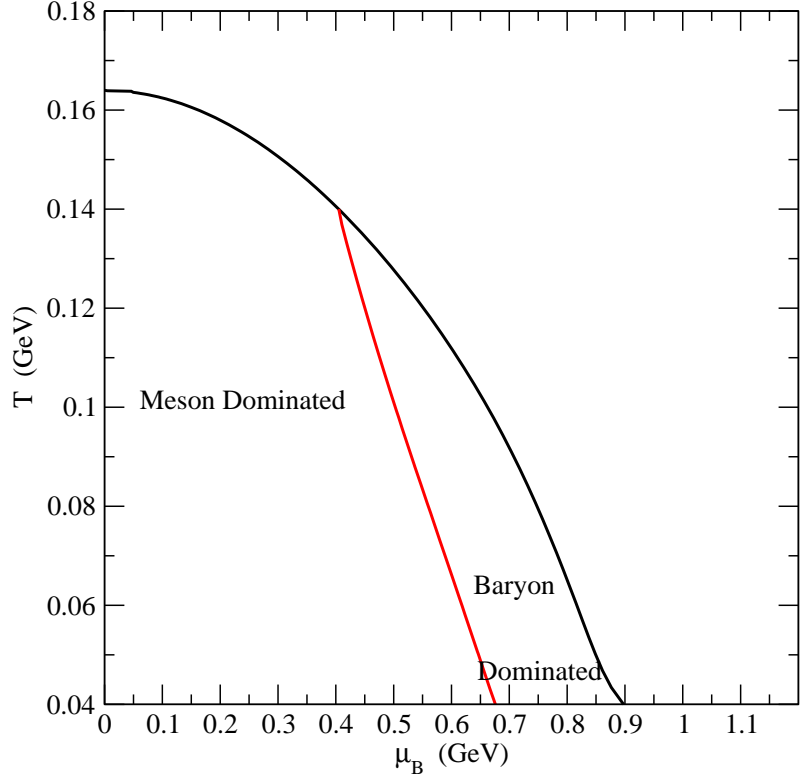}
 \caption{The line separating the $T-\mu_B$ plane into an area dominated by baryonic and one by mesonic freeze out
  as calculated in the Statistical
Model using {\sc Thermus}~\cite{thermus}.} \end{center}
\label{Mes_Bar_T_muB}
\end{figure}

\section{Deviations from the freeze-out curve}

In the previous section we have argued that the Statistical Model
with unique freeze out for all particles can not fully quantify
the sharpness of the K$^+/\pi^+$ ratio.   In this section, we
explore the possibility that freeze-out  might happen earlier in
this transition region.
For this interpretation, we show in Fig.~6 the calculated values
of the K$^+/\pi^+$ ratio for various combinations of $T$ and
$\mu_B$ as contour lines with the corresponding values given in
the figure. The thick solid line reflects the locations of the
freeze out given by the condition of Ref.~\cite{1gev}. If freeze
out happens around an incident energies of 30 $A$ GeV at higher
$T$, then the ratio K$^+/\pi^+$ will be higher. This ratio can
never exceed a value of 0.25 in an equilibrium condition.
\begin{figure}[h]
\begin{center}
\includegraphics[width=9.3cm]{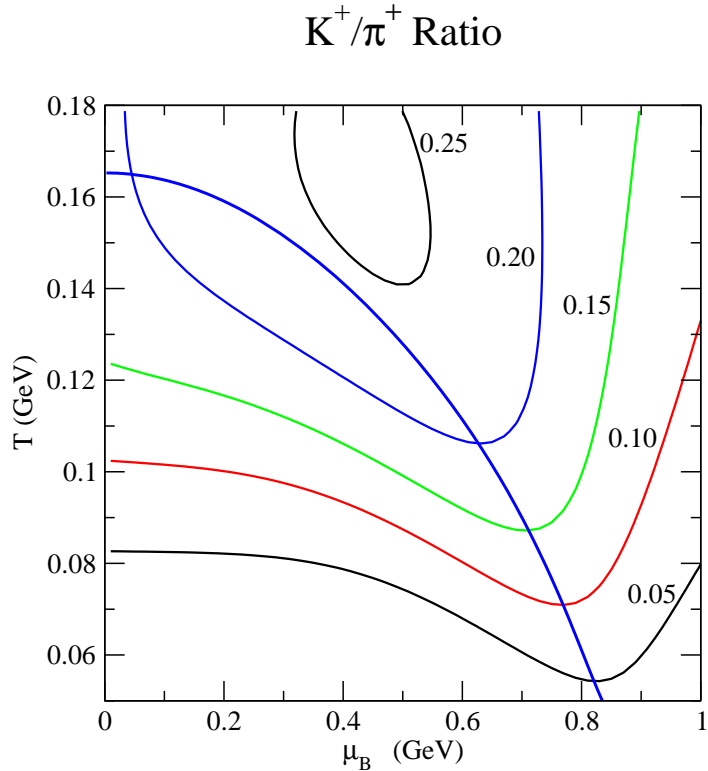}
 \caption{Values of the K$^+/\pi^+$ ratio for combinations
 of $T$ and $\mu_B$ are given by the contour lines and the corresponding values.
 The thick line refers to the freeze-out curve~\cite{1gev}.} \end{center}
\label{KP_PIP_T_mub}
\end{figure}

It turns out that other particle ratios are less affected by a
different freeze-out scenario, as their variation in the $T-\mu_B$
plane is very different~\cite{SW}.

\section{Other observations in this transition regime}

\begin{figure}[h]
\begin{center}
\includegraphics[width=9cm]{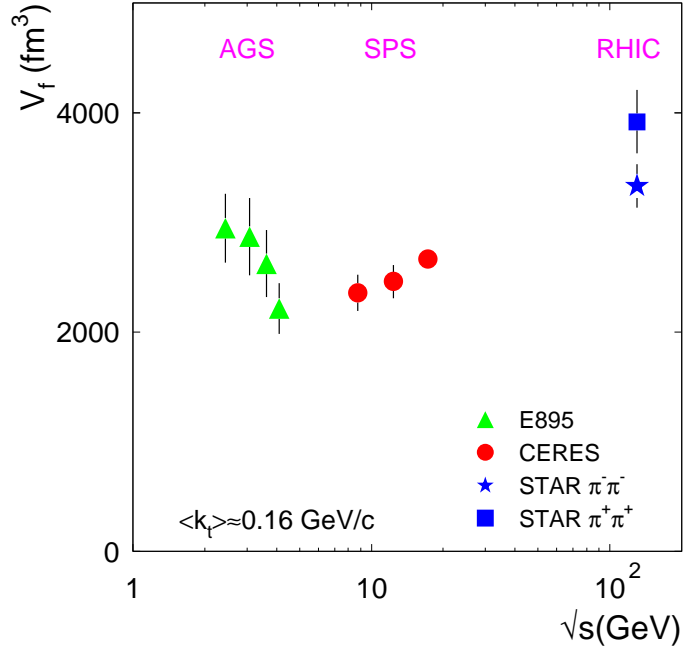}
 \caption{Freeze-out volume as extracted from HBT studies~\cite{CERES}.}
 \end{center} \label{CERES}
\end{figure}

An early freeze-out is also supported by results from HBT
studies~\cite{CERES}. Figure~7 shows the extracted volume as a
function of $\sqrt{s_{\rm NN}}$. Between top AGS and the lowest
SPS energies a minimum can be seen. As the fireball is expanding,
a smaller volume reflects an earlier time. The authors of
Ref.~\cite{CERES} relate this minimum to a change in the
interaction from $\pi N$ to $\pi \pi$. Indeed, assuming a mean
free path length of about 1 fm nicely explains the observed
trends. These studies have been continued and combined with the
volume extracted from the statistical model
fits~\cite{Andronic:2005yp,Andronic:2009qf}  and they all exhibit
a change of sign in this energy regime.

Furthermore, the pion multiplicity per number of participating
nucleons in heavy ion collisions crosses the results from pp
collisions also in this energy regime.

\section{Summary}

It has been shown that the Statistical Model yields a maximum in
the relative strangeness content around 30 $A$ GeV. This is due to
a saturation in the temperature $T$ while the chemical potential
keeps decreasing with incident energy. Since the chemical
potential plays a key role, it is clear that baryons are strongly
affected. Indeed, all hyperon/$\pi$ ratios yield maxima. In
contrast, the K$^-/\pi^-$ ratio shows a continuously rising curve
as expected from the arguments above. The K$^+/\pi^+$ ratio,
however, exhibits a maximum, as K$^+$ mesons are sensitive to the
baryo-chemical potential due to their associate production with
hyperons occurring at the lower incident energies. The model
predicts that for different hyperon/$\pi$ ratios the maxima occur
at different energies. If experiments prove this, the case for a
phase transition is strongly weakened.

The energy regime around 30 $A$ GeV seems to have specific
properties. It is shown that the entropy production occurs below
this energy mainly via creation of baryons, while at the higher
incident energies meson production dominates.

Finally, we put attention  on the impact of a change in the
freeze-out condition which might lead to an early freeze-out, thus
deviating from the usual freeze-out condition. Such a scenario
would increase the K$^+/\pi^+$ while leaving other particle ratios
essentially unchanged. HBT studies show that around 30 $A$ GeV a
minimum in the extracted volumina occurs. This could be
interpreted as an earlier kinetic freeze-out and might indicate
also another freeze-out for chemical decoupling.\\

This work was supported by the German Ministerium f\"ur Bildung
und Forschung (BMBF) and by the Polish Ministry of Science (MEN).


\begin{thebibliography}{99}
\bibitem{NA49} S.V. Afanasiev et al., (NA49 Collaboration),
Phys. Rev. {\bf C66} (2002) 054902. 
\bibitem{Gazdzicki}
M.~Ga\'zdzicki, (NA49 Collaboration), J. Phys. G: Nucl. Part.
Phys. {\bf 30} (2004) S701.
\bibitem{Lambda-NA49} T. Anticic et al.,
Phys. Rev. Lett. {\bf 93} (2004) 022302; C. Blume et al., J. Phys.
G: Nucl. Part. Phys. {\bf 31} (2005) S685.
\bibitem{:2007fe}
 {}C.~Alt {\it et al.}  [NA49 Collaboration]
 {}Phys.\ Rev.\  C {\bf 77}, 024903 (2008)
  [arXiv:0710.0118 [nucl-ex]]
\bibitem{:2008vb} 
 {}T.~Anticic {\it et al.}  [NA49 Collaboration]
 {}Phys.\ Rev.\  C {\bf 79}, 044904 (2009)
  [arXiv:0810.5580 [nucl-ex]]
\bibitem{Gorenstein}
M. Ga\'zdzicki and M.I. Gorenstein, Acta Phys. Polonica B {\bf 30}
(1999) 2705.
\bibitem{Stock} R.~Stock, J. Phys. G: Nucl. Part. Phys. {\bf 30} (2004) S633.
\bibitem{max_strange}  P. Braun-Munzinger, J. Cleymans, H.~Oeschler, K. Redlich,
Nucl. Phys. A 697 (2002) 902.
\bibitem{pi-AGS} L.~Ahle et al., (E802 Collaboration),
Phys. Rev. {\bf C57} (1998) 466.
\bibitem{Ahle_1999} L.~Ahle et al., (E802 Collaboration),
Phys. Rev. {\bf C60} (1999) 044904 and 064901.
\bibitem{Ahle_2000} L.~Ahle et al., (E866/E917 Collaboration),
Phys. Lett. {\bf B490} (2000) 53.
\bibitem{Lambda-AGS} S. Albergo et al., Phys. Rev. Lett. {\bf 88}
(2002) 062301.
\bibitem{Ahmad} S. Ahmad et al., Phys. Lett. {\bf B381} (1996) 3.
\bibitem{Klay}J.~Klay et al., (E895 Collaboration),
 Phys. Rev. {\bf C68} (2003) 054905.
\bibitem{Ahle} L.~Ahle et al., (E-802 Collaboration), Phys. Rev. {\bf C58}
(1998) 3523; Phys. Rev.{ C60} (1999) 044904; L.~Ahle et al.,
E866/E917 Collaboration, Phys. Lett.~{ B476} (2000) 1; Phys.Lett.
B490 (2000) 53.
\bibitem{Dunlop} J.C. Dunlop and C.A. Ogilvie, Phys. Rev. C61 (2000) 031901
and references therein; C. A. Ogilvie. Talk presented at QM2001,
Stony Brook, January 2001, Nucl. Phys. A {\bf 698} 3c; J.C.
Dunlop, Ph.D.Thesis, MIT, 1999.
\bibitem{bearden} I. Bearden et al., (NA44 Collaboration), Phys. Lett. B471
(1999) 6.
\bibitem{Star} J. Harris,
(STAR Collaboration), Talk presented at QM2001, Stony Brook,
January 2001, Nucl. Phys. A{\bf 698} 64c.

\bibitem{CLE99} J. Cleymans, H. Oeschler, K. Redlich, Phys. Rev. {\bf C59} (1999)
1663.
\bibitem{1gev} J. Cleymans and K. Redlich, Phys. Rev. Lett. {\bf 81}
(1998) 5284.
\bibitem{Andronic:2008gu}
  A.~Andronic, P.~Braun-Munzinger and J.~Stachel,
  Phys.\ Lett.\  B {\bf 673}, 142 (2009)
  [arXiv:0812.1186 [nucl-th]].
\bibitem{wroblewski} A. Wroblewski, Acta Physica Polonica~{B16} (1985) 379.
\bibitem{SK} S.~Kabana, Eur. Phys. J. {\bf C21} (2001) 545.
\bibitem{Cleymans:2004bf}
  J.~Cleymans, A.~Forster, H.~Oeschler, K.~Redlich and F.~Uhlig,
  Phys.\ Lett.\  B {\bf 603}, 146 (2004)
  [arXiv:hep-ph/0406108].
\bibitem{thermus} S. Wheaton, J. Cleymans and M. Hauer,
Computer Physics Communications, 180 (2009) 84-106.

\bibitem{Cle:2006}
J.~Cleymans, H.~Oeschler, K.~Redlich and S.~Wheaton, {}Phys.\
Rev.\  C {\bf 73}, 034905 (2006).
\bibitem{triplepoint} P.~Braun-Munzinger et al. to be published.
\bibitem{SW} S.~Wheaton, Ph.D.~thesis, University of Cape Town, 2005.
\bibitem{CERES} D. Adamova et al., (CERES Collaboration), Phys. Rev. Lett. {\bf 90}
(2003) 022301.
\bibitem{Andronic:2005yp}
  A.~Andronic, P.~Braun-Munzinger and J.~Stachel,
  Nucl.\ Phys.\  A {\bf 772}, 167 (2006)
  [arXiv:nucl-th/0511071].
\bibitem{Andronic:2009qf}
  A.~Andronic, P.~Braun-Munzinger and J.~Stachel,
  Acta Phys.\ Polon.\  B {\bf 40}, 1005 (2009)
  [arXiv:0901.2909 [nucl-th]].



\end{thebibliography}
\end{document}